\documentclass[prl,amsmath,amssymb,letter,preprint,endfloats]{revtex4}
\usepackage{graphicx}
\begin{document}
\title{Low relaxation rate in a low-Z alloy of iron}
\author{$^{1}$Christian Scheck, $^{1}$Lili Cheng, $^{2}$Igor Barsukov, $^{2}$Zdenek Frait, $^{1}$William E.
Bailey} \email{web54@columbia.edu (contact)} \affiliation{1. Dept.
of Applied Physics and Applied Mathematics, Columbia University, 500
W 120th St, New York, NY 10027, United States of America.\\2.
Institute of Physics, Academy of Sciences of the Czech Republic,
Prague, Czech Republic.}
\date{\today}
\pacs{75.25.+z,78.47.+p,76.50.+g,78.20.L.s,75.70.Ak}

\begin{abstract}

The longest relaxation time and sharpest frequency content in
ferromagnetic precession is determined by the intrinsic (Gilbert)
relaxation rate \emph{$G$}.  For many years, pure iron (Fe) has had
the lowest known value of $G=\textrm{57 Mhz}$ for all pure
ferromagnetic metals or binary alloys. We show that an epitaxial
iron alloy with vanadium (V) possesses values of $G$ which are
significantly reduced, to 35$\pm$5 Mhz at 27\% V.  The result can be
understood as the role of spin-orbit coupling in generating
relaxation, reduced through the atomic number $Z$.

\end{abstract}

\maketitle


Ultrafast magnetization dynamics comprise a major area of current
research in magnetism.  Novel dynamical phenomena have been observed
recently in confined structures\cite{back2}, with programmed field
pulses\cite{gerrits-nature,kaka-precessional}, through interactions
with intense light pulses\cite{koopmans-alpha,back4}, and under the
influence of spin polarized
currents\cite{rippard-prl-2004,ilya,heinrich-2003,kaka-locking}.

In all cases, the observed phenomena compete against ferromagnetic
relaxation in the magnetic material.  Relaxation aligns
magnetization $\mathbf{M}$ with applied fields $\mathbf{H}$,
bringing dynamics to a stop. The lowest limit of the relaxation rate
is intrinsic to a given material and given by $G=\gamma\alpha
M_{s}$, where $\alpha$ is the related dimensionless damping
constant.  In metals, the damping has seen renewed theoretical
interest\cite{kambersky-prb,steiauf} motivated particularly by its
formal relationship with spin momentum transfer
torques\cite{rebei-prb,choi-radiation,tserk-rmp}, or by its
enhancement with impurities\cite{rebei-re,reidy-apl}.  Low
relaxation rates are of particular interest for low critical
currents in spin momentum transfer excitation\cite{slonc},
narrowband response in magnetic frequency domain
devices\cite{scheck-apl}, and reduced thermal noise in nanoscale
magnetoresistive sensors\cite{neilsmith-snr}.

Pure iron (Fe) has long been known to exhibit the lowest measured
intrinsic relaxation rate of all elemental ferromagnetic metals or
binary alloys\cite{l-b-lambda-trunc}. Lowest values of 57 Mhz
($\alpha=\textrm{0.002}$) have been found in both single-crystal
whiskers and epitaxial films\cite{gilbert-g,frait-fraitova} at room
temperature. Elemental Ni, Co, and standard alloys such as
Ni$_{81}$Fe$_{19}$ show much higher values ($G=$220 Mhz, 170 Mhz,
114$\pm$10 Mhz, respectively.)

In this Letter, we show that the intrinsic relaxation rate $G$ in a
low-$Z$ ferromagnetic alloy can be substantially lower than that
known for pure Fe. Epitaxial MgO(100)/Fe$_{1-x}$V$_{x}$(8 nm)(100)
ultrathin films, deposited by UHV sputtering, exhibit values of $G$
to 35 Mhz, reduced by some 40\%.  While a comparable value has been
identified recently in NiMnSb\cite{heinrich-heusler}, the low
damping has been attributed to the special electronic
characteristics of this ordered, {\it half-}metallic
compound\cite{degroot-hm}, including a very low orbital component of
the magnetic moment.  We show that in Fe$_{1-x}$V$_{x}$ the observed
effect can be understood instead as the reduced influence of
spin-orbit coupling in lighter ferromagnets, pervasive across the
$3d$ series.


The intrinsic (or Gilbert) relaxation rate \emph{$G$} is defined in
the Landau-Lifshitz-Gilbert (LLG) equation for magnetization
dynamics (cgs units):
\begin{equation}
\frac{\partial\mathbf{M}}{\partial t} =
-\mid\gamma\mid\mathbf{M}\times\mathbf{H_{eff}}+\frac{G}{\gamma
M_{s}^{2}}\mathbf{M}\times\mathbf{\frac{\partial M}{\partial
t}}\label{llg}
\end{equation}

where
$\mathbf{H_{eff}}=\mathbf{H_{ext}}+\mathbf{H_{K}}+\mathbf{H_{demag}}$
is the effective field with external, anisotropy, and
demagnetization components and $\gamma$ = $g_{eff}$(e/2mc) =
($g_{eff}$/2)$\times$17.588 Mhz/Oe is the gyromagnetic ratio, with
$g_{eff}$ the spectroscopic factor.  Equation (\ref{llg}) can be
solved for power absorption in transverse RF susceptibility, as
performed in a ferromagnetic resonance (FMR) measurement. Damping
$\alpha$ is measured through variable-frequency FMR linewidth as

\begin{equation}
\Delta H_{pp}=\Delta H_{0}+ {2\over \sqrt{3}}{\alpha\omega\over
\gamma}\label{lw}
\end{equation}

where $\Delta H_{pp}$ is the field-swept FMR peak-to-peak linewidth,
$\Delta H_{0}$ is the inhomogeneous (extrinsic) broadening, and
$\omega/2\pi$ is the microwave frequency.  Extrinsic losses are
thought to be entirely microstructure-related, and (in principle)
possible to reduce through optimized microstructure.  $G$ is derived
from measurements of $\alpha$ through $G=\alpha\gamma M_{s}$.

In our experiments, $\alpha$ is measured in variable frequency,
field-swept FMR using four separate shorted rectangular waveguide
assemblies at 17, 25, 49, and 70 Ghz, with 2 mm iris, see
\cite{gilbert-g} for details. Frequency-swept FMR measurements were
carried out using a synthesized microwave sweep generator operating
in CW mode over the range 9-40 Ghz\cite{scheck-apl}, with sample
mounted on a broadband coplanar waveguide (CPW), over an exciting
area of 0.1 mm$^{2}$; an identical setup was used for the
identification of field-for-resonance $\omega(H_{res})$.  All
measurements were carried out at room temperature, with
magnetization in the film plane, along
[110]Fe$_{1-x}$V$_{x}$/[100]MgO unless otherwise noted.

Two series of epitaxial MgO(100)/Fe$_{1-x}$V$_{x}$(100) samples have
been considered.  A 50nm series, to 52\% V, was investigated for
magnetic moment $4\pi M_{s}$ by vibrating sample magnetometry (VSM),
magnetocrystalline anisotropy constant $K_{2}$ by
rotation-controlled FMR in a X-band cavity, and spectroscopic factor
$g_{eff}$ by $\omega(H_{res})$ measurement.  8nm samples have been
compared between pure Fe and 27\% V for Gilbert relaxation rate $G$;
ultrathin samples are necessary to exclude eddy current effects.
This series had optimized, high rate deposition conditions for
lowest $G$ in pure Fe; the V composition of 27\%, shown here, was
the maximum attainable in our chamber for these conditions.


\begin{figure}[htb]
\includegraphics[width=\columnwidth]{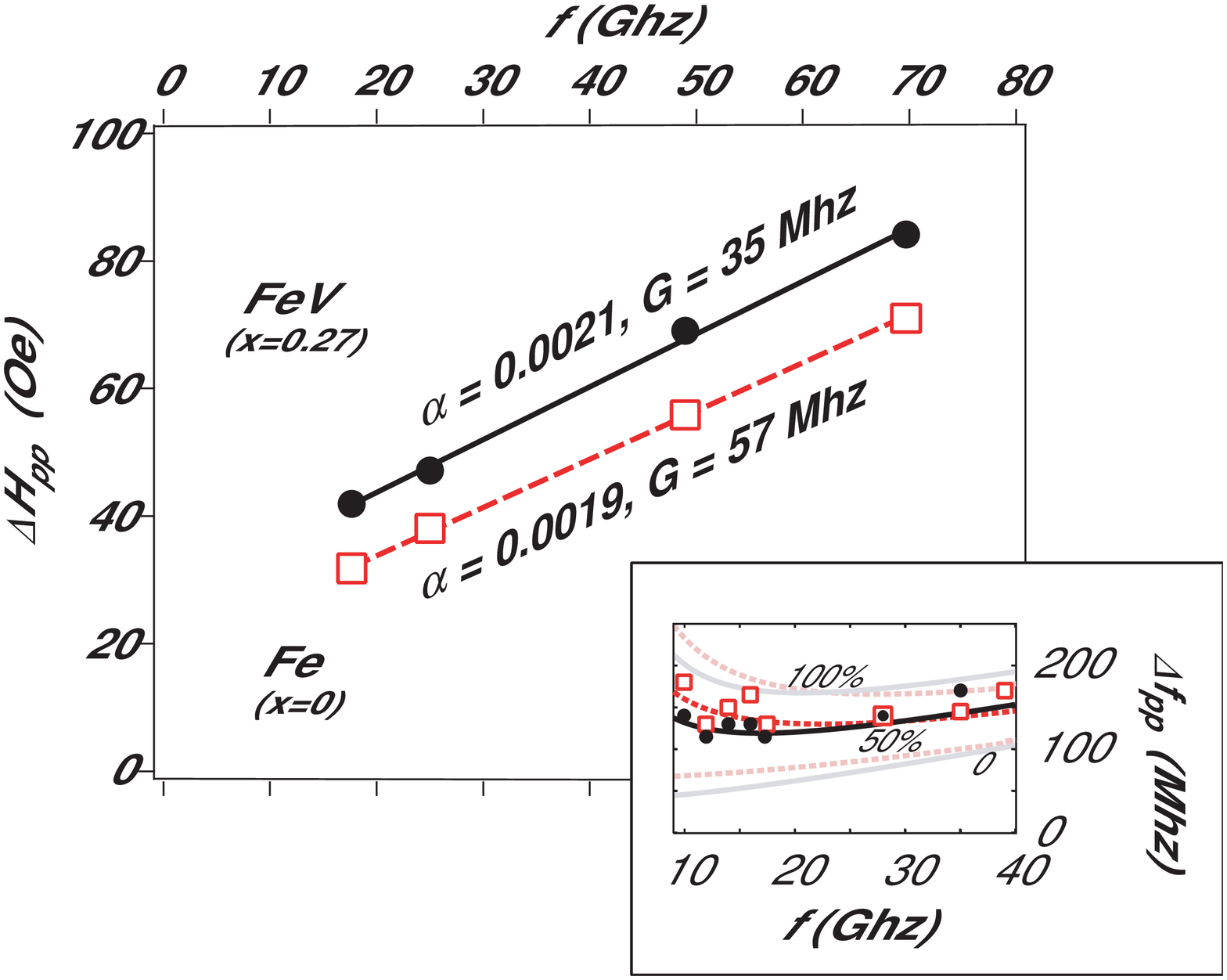}
\caption{Measurement of intrinsic damping parameter $\alpha$ and
relaxation rate $G$ in MgO/Fe$_{1-x}$V$_{x}$(8 nm), $x=0$ and
$x=\textrm{0.27}$. {\it Inset:} swept-frequency FMR linewidths
$\Delta f$; lines show model calculations for different levels of
inhomogeneous broadening.\label{frait-data}}
\end{figure}

Figure \ref{frait-data} shows the variable-frequency, field-swept
FMR data for 8 nm Fe$_{1-x}$V$_{x}$ alloy films, $x= 0$ and
$x=\textrm{0.27}$.   We have determined $\alpha$ through a linear
fit of $\Delta H(\omega)$ according to equation (\ref{lw}).  For
pure Fe, $\alpha$ is measured as 0.0019, in good agreement with the
lowest measured values through exchange-conductivity analysis of
single-crystal whiskers and sputtered epitaxial films with higher
inhomogeneous loss\cite{gilbert-g}; the inhomogeneous loss $\Delta
H_{0}$ is 18 Oe.  For Fe$_{1-x}$V$_{x}$, $x=\textrm{0.27}$, we
measure $\alpha = \textrm{0.0021}$, with $\Delta H_{0}=\textrm{28
Oe}$.

Values of $4\pi M_{s} = \textrm{21.1 kG}$ and 11.6 kG, are measured
for these 8nm films, respectively, through perpendicular resonance
(not shown) . Calibrated VSM measurements (Figure \ref{f2}) for
thicker (50 nm) films return values which are $\sim$ 0.5 kG higher
in each case, presumably due to some reduced moment at the surface
for the thinner films. The observed reduction in the moment for the
Fe$_{73}$V$_{27}$ alloy ($\sim$3.6 $\mu_{B}/\textrm{atom V}$) is
close to the theoretical moment reduction from V impurities in
Fe\cite{drittler} ($\sim$3.4 $\mu_{B}/\textrm{atom V}$) and the
Slater-Pauling value of 3.3 $\mu_{B}/\textrm{atom}$.

The intrinsic relaxation rate $G$, combining experimental $\alpha$
and $4\pi M_{s}$ measurements of the thin films, is 57 $\pm$ 3 Mhz
for Fe and 35 $\pm$ 5 Mhz for Fe$_{1-x}$V$_{x}$, $x=\textrm{0.27}$.
The reduction of $G$ has a device-relevant manifestation. FMR
frequency linewidths $\Delta f$ at half power (peak-to-peak) measure
$2G$ ($2/\sqrt{3}G$) directly, in the low-frequency, intrinsic
damping limit.  Swept-frequency linewidths $\Delta f_{pp}$ show
lower values for the 27\% V film by 15-30 Mhz over the frequency
range 10-18 Ghz, as expected in analytical calculations assuming
zero, 50\%, and 100\% of the swept-field measured inhomogeneous
loss; calculations convert the field-swept linewidth $\Delta H$ to
$\Delta \omega$ by differentiating the Kittel relationship.
Agreement is best for 50\% (dark curve in Fig. 1); the reduced value
of inhomogeneous broadening presumably reflects the smaller sampling
area of the broadband technique.

\begin{figure}[htb]
\includegraphics[width=\columnwidth]{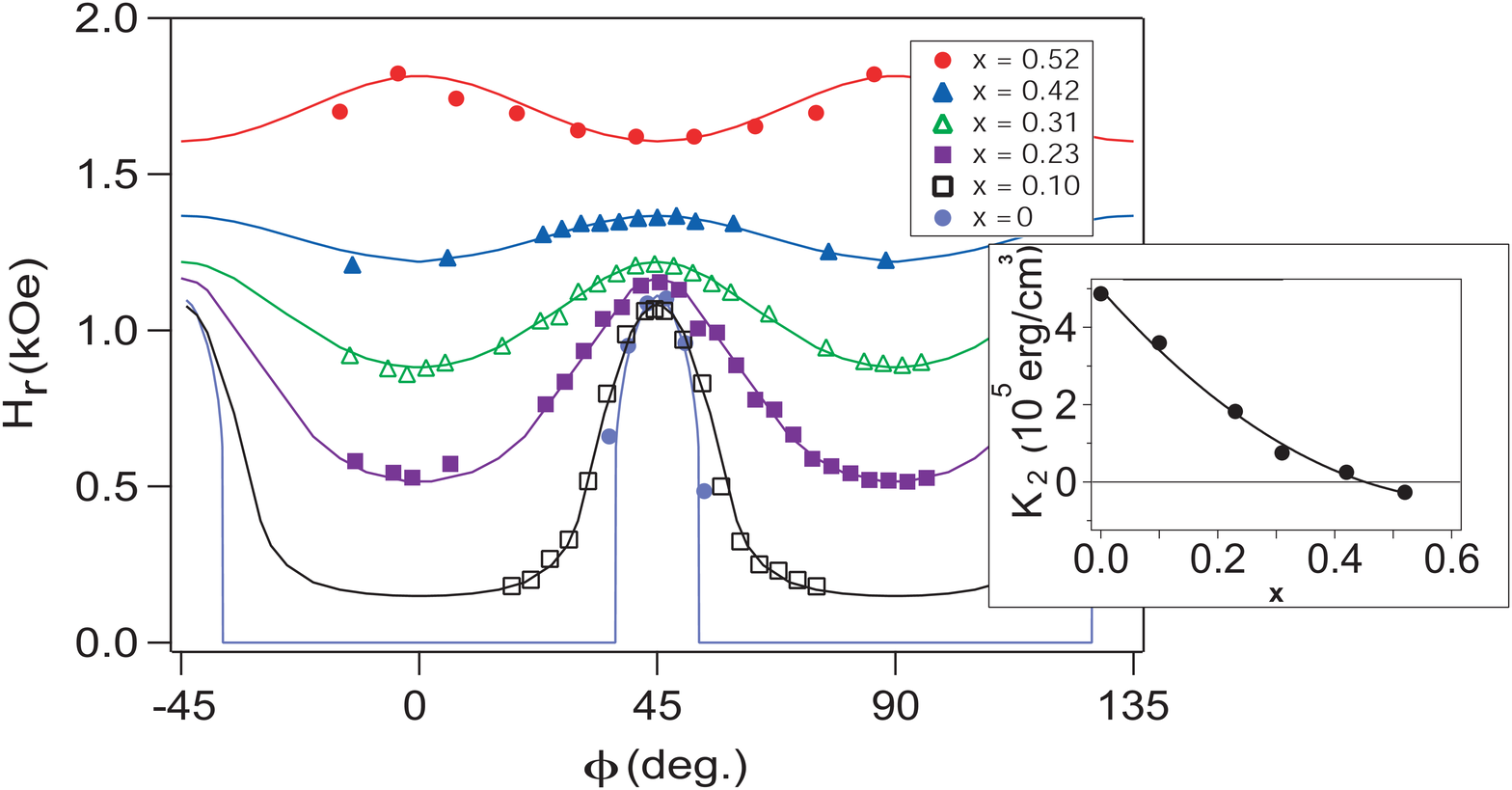}
\caption{Angular dependence of field-for-resonance H$_{res}$ at 10
Ghz, cavity measurement.  Lines are fits to extract cubic anisotropy
constants $K_{2}$ after \cite{farle-rpp}.  {\it Inset:}
\emph{$K_{2}$} values extracted from fits.\label{K4}} \label{af}
\end{figure}

We have also characterized the magnetocrystalline anisotropy in 50
nm Fe$_{1-x}$V$_{x}$ film series (Fig. \ref{af}.)  The angular
dependence of $H_{res}$ exhibits a clear fourfold symmetry with
minima along [100] and [0$\overline{1}$0] (easy axes) and maxima
along [110] and [1$\overline{1}$0] (hard axes).  As V concentration
increases, the magnetocrystalline anisotropy is reduced to zero and
negative values at $x\sim\textrm{0.44}$ (inset).  Numerical fits to
extract the cubic magnetocrystalline anisotropy constant $K_{2}$
were performed according to the method of Ref. \cite{farle-rpp}.
Extracted values yield the expected value of 4.87$\times$10$^{5}$
erg/cm$^{3}$ for Fe; we find 1.29$\times$10$^{5}$ erg/cm$^{3}$ for
Fe$_{73}$V$_{27}$.  While data for $\textrm{0}\leq x \leq
\textrm{10\%}$ match closely with the findings of
Hall\cite{hall-fe}, who investigated to 15\%, the nulling of $K_{2}$
at high $x$ had not been identified previously.

The use of variable frequency FMR measurement, with knowledge of
$4\pi M_{s}$ values, allows the extraction of the gyromagnetic ratio
$g_{eff}$, through the Kittel relation,
$\omega=\gamma\sqrt{\left(H_{B}+4\pi
M_{s}+K_{2}/M_{s}\right)\left(H_{B}-2K_{2}/M_{s}\right)}$,
appropriate for film magnetization in-plane along <110>. A plot of
$f^{2}$ as a function of $H_{r}$, extracted from variable frequency,
field-swept FMR measurement of 50 nm films, is presented in Fig.
\ref{f2}. Values of $4\pi M_{s}$, taken from VSM measurements, were
fixed in the fits; $g_{eff}$ and $K_{2}$ were fit parameters.
Extracted \emph{$K_{2}$} values are in good agreement with those
taken from rotational measurements.

Little change in $g_{eff}$ is evident as the V concentration
increases.  \emph{g$_{eff}$} values, shown in the inset of Fig.
\ref{f2}, are near 2.09 $\pm$ 0.02 for V composition up to 52\%.

\begin{figure}[htb]
\includegraphics[width=\columnwidth]{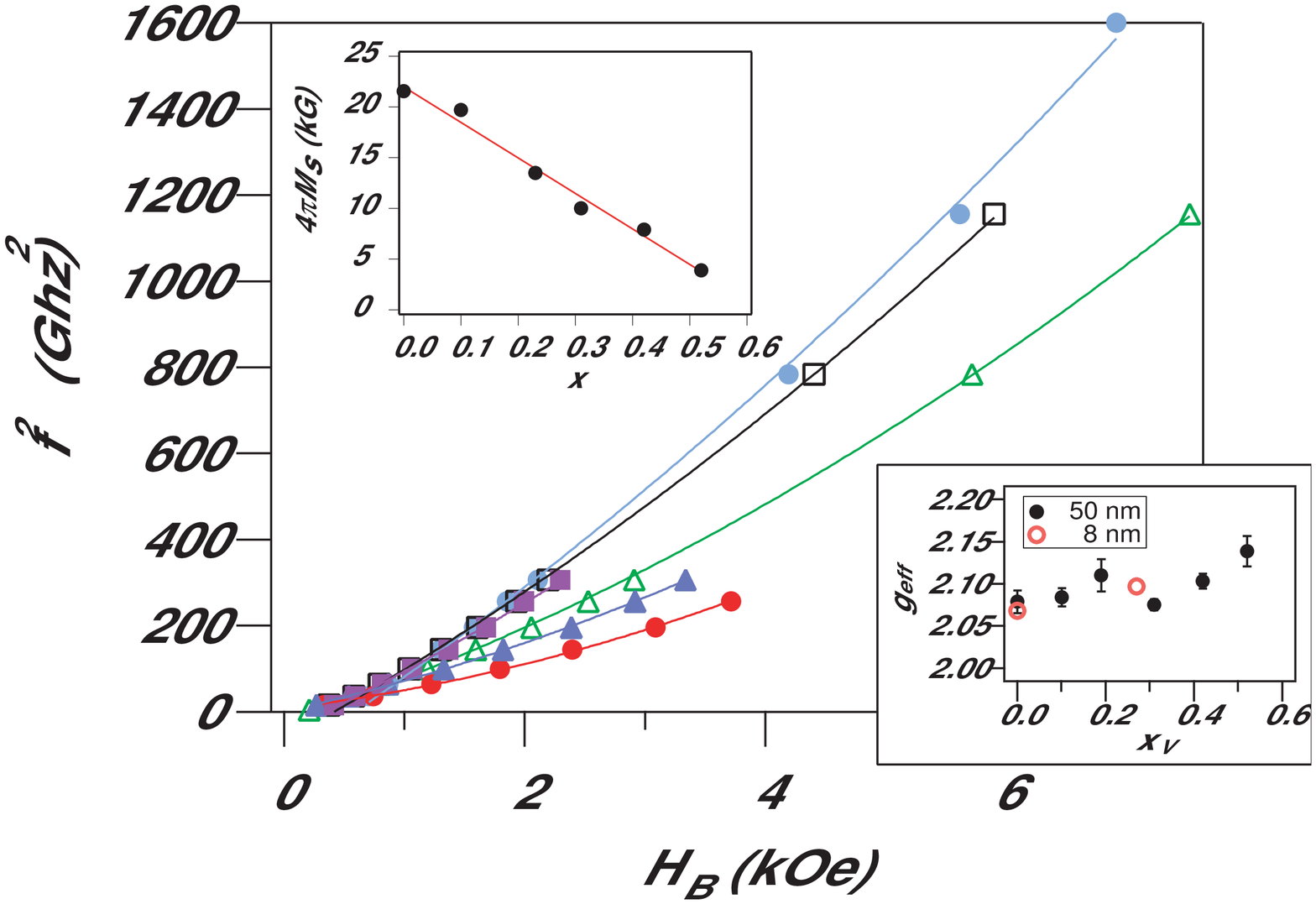}
\caption{Extraction of spectroscopic factor $g_{eff}$ from
variable-frequency FMR measurement, MgO/Fe$_{1-x}$V$_{x}$(8 and 50
nm) films.  {\it Inset, top left:} VSM measurement of $4\pi M_{s}$.
{\it Inset, lower right:} extracted \emph{g$_{eff}$} values for 8 nm
and 50 nm films.\label{f2}}
\end{figure}

\begin{figure}[htb]
\includegraphics[width=\columnwidth]{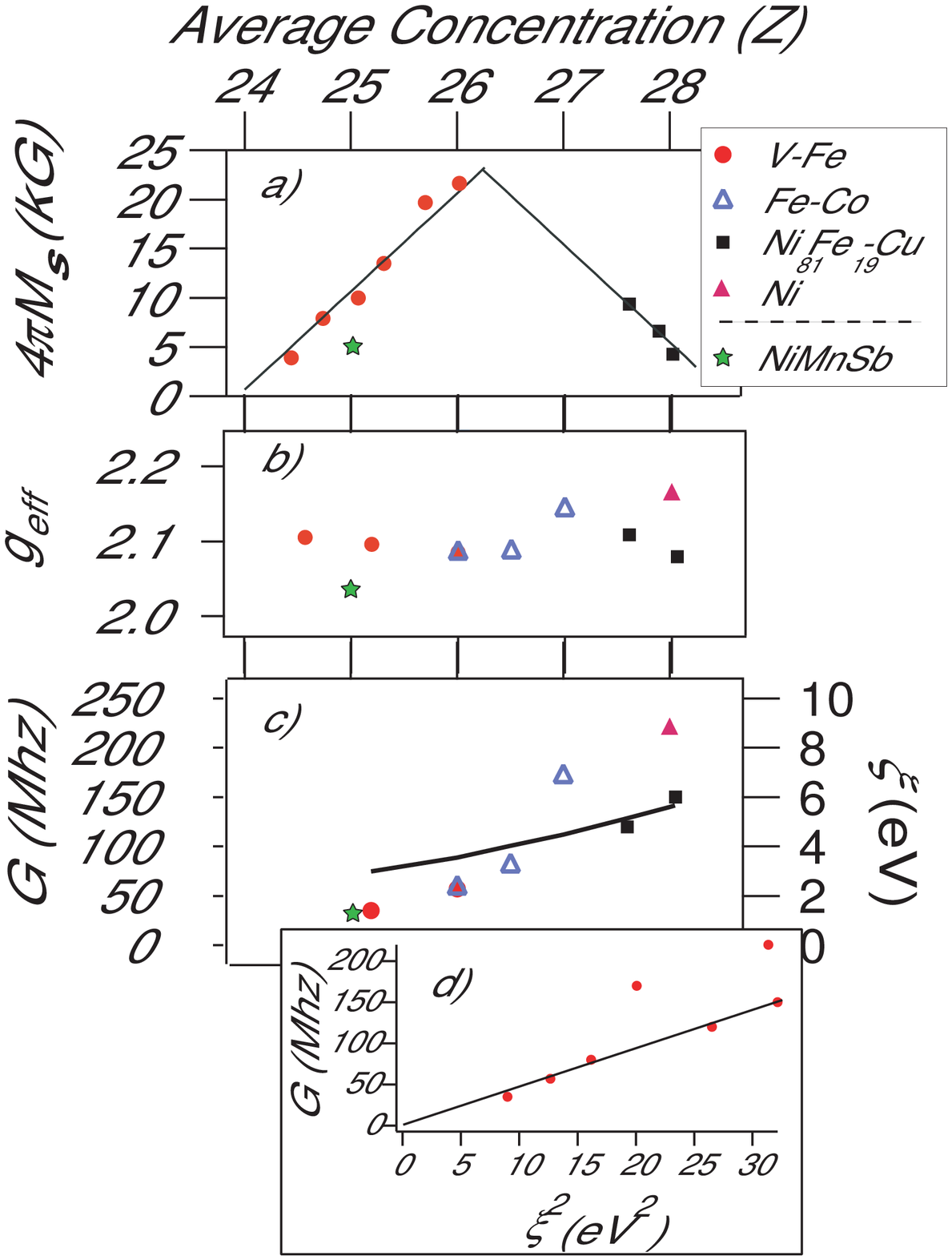}
\caption{Compositional dependence of a) moment $4\pi M_{s}$, b)
gyromagnetic ratio $g_{efF}$, and c) Gilbert relaxation rate $G$ for
3d transition metal ferromagnetic substitutional alloys.  Solid
line: atomic spin-orbit coupling parameter $\xi$.  d): Relaxation
$G$ as a function of spin orbit coupling $\xi^{2}$.  Data for Fe-V
films, this work, for BCC/FCC Fe$_{1-x}$Co$_{x}$, ref
\cite{gilbert-g}; for Ni$_{81}$Fe$_{19}$:Cu films, \cite{guan-pycu};
for Ni whiskers\label{allcomp}, \cite{bhagat-lubitz}.  The Heusler
compound NiMnSb data\cite{heinrich-heusler} are included for
comparison.}
\end{figure}

We have shown that Fe$_{1-x}$V$_{x}$ alloys exhibit a reduction in
intrinsic relaxation rate $G$.  The magnetic moments of this alloy
have been understood previously through a simple dependence on
average electronic concentration $Z$, as seen in the well-known
Slater Pauling curve\cite{staunton-jap}.  The VSM measurements of 50
nm films are presented together with the theoretical result in
Figure \ref{allcomp}, {\it top}.  $4\pi M_{s}$ data are also
included from perpendicular FMR measurements of Cu-alloyed
Ni$_{81}$Fe$_{19}$, at $27.6\leq Z\leq 28.0$, taken from ref.
\cite{guan-pycu}.  Heusler alloy data, for
NiMnSb\cite{heinrich-heusler} are included for comparison and
plotted at $Z=25$, as magnetic properties are thought to arise
nearly entirely from moments localized to Mn sites\cite{degroot-hm}.

In Fig \ref{allcomp}c) we plot measurements of intrinsic relaxation
rate $G$ as a function of average electronic concentration $Z$. Data
are taken from our measurements and collected from the literature.
For consistency, we restrict ourselves to measurements of $G$ in
crystalline films, using variable frequency FMR, in which the
frequency-dependent change in (homogeneous) linewidth
$\textrm{1.158}\Delta \omega\alpha/\gamma$ exceeds the inhomogeneous
linewidth $\Delta H_{0}$ by at least a factor of two. This requires
70 Ghz measurements for Fe$_{1-x}$V$_{x}$ (this work) and
Fe$_{1-x}$Co$_{x}$\cite{gilbert-g} and 18 Ghz measurements for
Ni$_{81}$Fe$_{19}$ and its alloys\cite{guan-pycu}.  Lowest
literature values are plotted; in e.g. Fe, a dispersion of measured
values from 57-140 Mhz\cite{lubitz-jap-cu-fe,heinrichPRL03} has been
attributed to variations in point defect density\cite{safonov-jap}.

The data in Fig \ref{allcomp}c), demonstrate a trend towards higher
relaxation rate $G$ at higher average valence $Z$.  Values increase
by a factor of four, from 35 Mhz for FeV(27\%) to 150 Mhz - 220 Mhz
for (Ni$_{81}$Fe$_{19}$)$_{0.7}$Cu$_{0.3}$ and for pure Ni,
respectively. Apart from the two pure metal points for Ni whiskers
and the FCC Co film, there is a relatively smooth variation in the
baseline formed by the alloys.  For these species, there is no
similar general trend in $g_{eff}$ ({\it middle panel}), measured
constant at $\textrm{2.10}\pm\textrm{0.02}$.  Note that the NiMnSb
point, of $G=\textrm{31 Mhz}$ at $Z=25$, is very close to our
measurement of Fe$_{73}$V$_{27}$ of 35 $\pm$ 5 Mhz, even though
$g_{eff}$ is substantially lower (2.03 compared with 2.11).

As average concentration $Z$ increases, so too does the expected
spin orbit coupling energy.  Values for an effective atomic
spin-orbit coupling parameter $\xi$, where
$H_{s.o.}=\xi\sum_{i}\mathbf{l}_{i}\mathbf{s}_{i}$, have been
tabulated in Ref. \cite{d-so-calc} for atomic $3d$ orbitals, in good
agreement with atomic spectra. These values are reproduced here (Fig
\ref{allcomp}c), solid line). In Figure \ref{allcomp}d), we show the
dependence of $G$ on $\xi$, implicit in atomic number. It can be
seen that the lower band of values is simply proportional to
$\xi^{2}$; pure Ni and Co data points are significant outliers.

The observed scaling of Gilbert damping $G$ with $\xi^{2}$ is in
good agreement with electronic-scattering based models of
ferromagnetic relaxation, appropriate to
metals\cite{kambersky-microscopic}.  Relaxation occurs as
uniform-mode magnons are annihilated by one-electron spin-flip
accelerations.  The expression for relaxation rate $G$ is given
as\cite{heinrich-ieee,heinrich-review}

\begin{eqnarray*}
G = \hbar\gamma^{2}<S>^{2}\xi^{2}\int d^{3}\mathbf{k}
\sum_{\alpha,\beta,\sigma}<\beta|L^{+}|\alpha><\alpha|L^{-}|\beta>\cdot\\
\times\delta(E_{\alpha,\mathbf{k},\sigma}-E_{F})\cdot{\hbar/\tau_{M}\over
\left(\hbar\omega+E_{\alpha,\mathbf{k},\sigma}-E_{\beta,\mathbf{k+q},\sigma}\right)^{2}+\left(\hbar/\tau_{M}\right)^{2}}
\end{eqnarray*}

where other leading materials parameters $\gamma^{2}$ and
$<S>^{2}=(M(300K)/M(0K))^{2}$ do not change by more than 5\% or
10\%, respectively, across the materials investigated.  The second
row terms describe the temperature dependence of relaxation through
momentum scattering $\tau_{M}$, with a high-temperature limit
$G\sim\tau_{M}^{-1}\sim T$, for scattering across spin sheets
("interband scattering") and a low-temperature limit
$G\sim\tau_{M}\sim 1/T$ for scattering within bands ("ordinary
scattering.")  The temperature dependence of $G$ is known to be weak
near 300K for Fe, Co, Ni\cite{bhagat-lubitz} and
Ni$_{81}$Fe$_{19}$\cite{bailey-Tb}, although it is not known with
great precision except for
Ni.\cite{heinrich-fmar-ni,heinrich-cochran-fmar-fe} Thus it is not a
gross distortion to take these terms as nearly constant, near 300K,
across the $3d$ series.  The orbital moment terms, on the other
hand, may not be constant: orbital moments do {\it not} follow a
simple dependence upon $Z$ in binary Fe,Co,Ni
alloys\cite{meyer-asch-jap-1962}, and the significantly larger
values of $g_{eff}$ observed for Co ($g_{eff}=\textrm{2.15}$) and Ni
($g_{eff}=\textrm{2.17}$) may help to explain their significantly
higher values of $G$.

We do not expect that a universal proportionality between $G$ and
$\xi^{2}$--a simple dependence upon $Z$--should exist. Deviations
are noted for Ni and Co, materials with large $g_{eff}$.
Ferromagnets with very large magnetostriction\cite{suhl} are likely
to be dominated by phonon drag mechanisms, weak in the materials
under consideration here\cite{mac-heinonen}.  At low temperature and
long scattering time $\tau_{M}$ the details of band structure,
variable across the series, should become much more
important\cite{kambersky-prb,steiauf}.  Nevertheless, the scaling of
relaxation rate $G$ with spin-orbit coupling as $\xi^{2}$ provides a
plausible interpretation for the very low relaxation rates seen in
the low-Z alloy Fe$_{1-x}$V$_{x}$.  It explains why similarly low
values are seen in the very different Heusler compound NiMnSb.
Finally, we hope that it will provide some guidance in the search
for ferromagnetic materials with even lower relaxation rates.


In summary, we report on the discovery of an alloy with the lowest
intrinsic relaxation rate, $G\sim\textrm{35}\pm\textrm{5 Mhz}$ yet
observed in ferromagnetic metals.  The results meet a critical need
in metallic ferromagnetic materials, helping to enable high-$Q$
integrated microwave devices, and fostering emerging devices based
on spin momentum transfer (SMT)\cite{ilya,kaka-locking} excitations.


We thank O. Myrasov, M. Fahnle, M. Stiles, A. Rebei, L. Berger, and
B. Heinrich for discussions.  This work was supported by the US Army
Research Office under contracts DA-ARO-W911NF0410168,
DAAD19-02-1-0375, and 43986-MS-YIP, and has used the shared
experimental facilities that are supported primarily by the MRSEC
program of the National Science Foundation under NSF-DMR-0213574.

\end{document}